\documentclass[10pt,twocolumn,prd,aps,amssymb,amsmath]{revtex4}
\usepackage{graphics}
\usepackage[pdftex]{graphicx}

\newcommand{\ket}{\rangle}
\newcommand{\bra}{\langle}

\begin{document}
\preprint{}

\title{Reexamination of the optimal Bayes cost in the binary decision problem}

\author{Bernhard K. Meister}
\email{b.meister@imperial.ac.uk}
\affiliation{ Department of Physics, Renmin University of China, Beijing, China 100872}

\date{\today }

\begin{abstract}

The problem of quantum state discrimination between two wave functions of a particle in a square well potential is considered. The optimal minimum-error probability 
is known to be given by the Helstrom bound. A new strategy is introduced by inserting  an impenetrable barrier in the middle of the square well, which is either a nodal or non-nodal point of the wave function. 
The energy required to insert the barrier is  dependent on the initial state. This enables the experimenter to gain additional information beyond the standard probing of the state envisaged by Helstrom and to improve the distinguishability  of the states. It is shown that under some conditions 
the Helstrom bound can be violated, i.e. the state discrimination can be realized with a smaller error probability.

\end{abstract}
\maketitle


\section{Introduction}
\label{sec:1a}

Experimental design and data analysis are common challenges in science, and particular acute in quantum mechanics, due to recurring questions about foundational issues and restrictions on measurements.
In the literature many different approaches are discussed. For example, there is the information theoretic approach, where one maximizes the mutual information, and the minimax approach, where one minimizes the maximum cost of a set of strategies assuming a perfidious opponent. Both methods are popular, but this paper will instead use the Bayes procedure to minimize the expected cost, since the existence of a {\it prior} associated with the states to be distinguished is assumed.
In an earlier paper \cite{bkm2010} the author described another approach based on isoenergetic compression, which also enables under special conditions a break of the well-known Helstrom bound.

 Two disparate concepts are combined in the paper: quantum state discrimination and the modification of the  quantum potential.
Quantum state discrimination or more generally called Bayesian hypothesis testing was developed by Helstrom and others \cite{helstrom, holevo, yuen}. The particular problem of quantum state discrimination  between two possible states with given prior and transition probability was also studied in various settings, e.g. for a more recent paper on this topic see Brody {\it et al.} \cite{dbbm}. It is generally accepted, but will be challenged in the paper, that the optimal Bayes cost in the binary case, is given by the Helstrom bound, which only depends on the prior and the transition probability between the states and can be written in a simple closed form. 
 An instantaneous insertion of a barrier into an one-dimensional infinite well was studied in Bender {\it et al.} \cite{bbm}. The authors established that this type insertion has various curious effects like creating new set of eigenstates on both sides of the impenetrable barrier with the associated shifts in the probabilities. Here we study implications for the distinguishability of states. 

Next a description of the setup and the procedure to be analyzed. With fixed probabilities, given by the prior, one of two quantum states is put into a square well with infinitely high walls.
Two strategies for calculating the Bayes cost are proposed. In the first strategy,  the combination of prior and transition probability between the two quantum states alone is sufficient to calculate the conventional optimal minimum error probability, i.e. the Helstrom bound. This
error probability can be achieved by a well-known measurement strategy \cite{helstrom} not further discussed in the paper.
  The second  strategy for calculating the error probability is novel. It starts with the instantaneous insertion of an impenetrable barrier into a square well. The energy required to insert a barrier
differs for each state.





There are two main motivations for the work presented here. On the one hand it will possibly shed some light on foundational issues in quantum measurement theory, and on the other hand there are practical problems in quantum information theory, which depend on optimal state discrimination.

The structure of the paper is as follows. In section two the impact of an insertion of a barrier in a one-dimensional infinite well is studied. In  the third section the binary choice problem between two quantum particles in this well is tackled.
In the conclusion the result is briefly restated  and some general comments added.





 \section{Insertion of a barrier into a one-dimensional well}
\label{sec:1b}
In this section the instantaneous insertion of impenetrable barriers into an infinite square well at both nodal or non-nodal points is presented.
 A particle of mass $M$ and with wave function $\phi(x)$ is trapped in a one-dimensional infinite square well of width $L$. The Hamiltonian is given by
\begin{eqnarray}
H= - \frac{\hbar^2}{2M}\frac{d^2}{dx^2}.\nonumber
\end{eqnarray}
At the boundary points, $x=0$ and $x=L$, and outside the interval the wave function  vanishes.
The eigenfunctions of the Hamiltonian satisfying the boundary conditions are
\begin{eqnarray}
\phi_n (x)= \sqrt{ \frac{2}{L}}\sin\Big(\frac{n \pi x }{L}\Big),\nonumber
\end{eqnarray}
with the respective energy eigenvalues $E_n(L)=\frac{\pi^2 \hbar^2 n^2}{2 M L^2}$.

In the following paragraphs we discuss the insertion of a barrier, which  can be divided into two cases. In the first case one inserts the barrier at
a nodal point, and in the second case the insertion happens at a point of non-zero amplitude.
In the first case the calculation is  easier, since the energy of the system is left unchanged.
Let us look at the example of the second excited state 
\begin{eqnarray}
\phi_{before}(x)=\sqrt{\frac{2}{L}}\sin\Big(\frac{2 \pi x} {L}\Big).\nonumber
\end{eqnarray}

If the insertion happens at $t=0$ at the nodal point $L/2$ where the wave function vanishes, the result is  the non-normalized first excited eigenstate of the smaller compartments on the left and right of the barrier
\begin{eqnarray}
\phi_{after}(x)
\left\{
\begin{array}{lr}
\sqrt{\frac{2}{L}}\sin( \pi 2x  / L), &(0\leq x \leq \frac{1}{2}L),\\\nonumber
\sqrt{\frac{2}{L}}\sin( \pi 2x  / L), & (\frac{1}{2}L\leq x \leq L).\\
\end{array}  \right. \nonumber
\end{eqnarray}
Significantly, the expectation value of the Hamiltonian before and after the insertion remains unchanged, since in the definition of the energy both the size of the well and the counter of the energy level appear as squares. This means the experimenter can carry out the procedure without adding or extracting energy.

Next we consider the more intricate case of an insertion at a non-nodal point. As an example we study the  
first excited state 
 \begin{eqnarray}
\psi_{before}(x)=\sqrt{\frac{2}{L}}\sin\Big(\frac{ \pi x} {L}\Big).\nonumber
\end{eqnarray}
In this case the point $L/2$ is not a nodal point and an instantaneous insertion has all kinds of curious consequences.
The wave function can be expanded on both sides of the inserted barrier into Fourier series, which are  constructed out of the eigenstates of the respective Hamiltonians with the expansion for the left compartment
\begin{eqnarray}
\psi^L_{after}(x,0)=\sum_{n=0}^{\infty} b_n^L \sqrt{ \frac{4}{L}}\sin\Big(\frac{2n \pi x }{L}\Big)  
\end{eqnarray}
and the right compartment
\begin{eqnarray}
\psi^R_{after}(x,0)=\sum_{n=0}^{\infty} b_n^R \sqrt{ \frac{4}{L}}\sin\Big(\frac{2n \pi (x-L/2 ) }{L}\Big). \nonumber
\end{eqnarray}
A calculation of the coefficients for the initial function 
$\sqrt{ \frac{2}{L}}\sin(\frac{ \pi x }{L})$ on the left hand side, i.e in the interval $[0,L/2]$, using the formula
\begin{eqnarray}
b_n^L=\int_0^{L/2} dx \sqrt{ \frac{2}{L}}\sin\Big(\frac{\pi x }{L}\Big)\sqrt{ \frac{4}{L}}\sin\Big(\frac{2 n \pi x }{L}\Big)\nonumber
\end{eqnarray}
  leads to
\begin{eqnarray}
b_n^L= (-1)^n \frac{4\sqrt{2} n}{\pi-4n^2 \pi} .\nonumber
\end{eqnarray}
The calculation for the right hand side due to symmetry reasons is similar in nature and 
results in
\begin{eqnarray}
b_n^R= - \frac{4\sqrt{2} n}{\pi-4n^2 \pi} .\nonumber
\end{eqnarray}
 Based on the argument fully laid out in section IV of \cite{bbm}, one can say that the Fourier expansion converges  pointwise everywhere except at the insertion point, i.e. the  Gibbs phenomenon. The energy expectation for the wave function at the insertion time and excluding the energy at $L/2$ is given by
\begin{eqnarray}
 \int_{0}^{L/2} (\psi^L_{after}(x,0))^*(x,0) H \psi_{after}^L(x,0) dx\nonumber\\
 + \int_{L/2}^L (\psi^R_{after}(x,0))^*(x,0) H \psi_{after}^R(x,0) dx.\nonumber
\end{eqnarray}
The series representation of the energy in the left compartment is a divergent sum
\begin{eqnarray}
\sum_{n=0}^\infty (b^L_n)^2 \frac{2\pi^2 \hbar^2 n^2}{ML^2},\nonumber
\end{eqnarray}
since $b^L_n\sim 1/n$ for large $n$. However, this divergence does not imply an infinite energy at the insertion time itself. This is due to the fact that the series defined in ($1$)  does not converge uniformly to $\psi^L_{after}(x,0)$. If one takes this into account and does the appropriate manipulations as shown in \cite{bbm}, the total energy of the left and right compartment stays unchanged.
The reader should not be mislead to believe that no energy is needed to introduce a barrier, since the insertion itself changes the energy at the point $L/2$. The energy needed in an idealized setting to insert the barrier is infinite. 
The energy localized in the barrier point propagates through the system at $t>0$ and increase the energy in both chambers.  The result is a fractal wave function similar to the type discussed in Berry \cite{berry}.
The system is still described as a whole by a single quantum state, which can be probed jointly or separately in each region.
To summarize, 
the energy needed to insert the barrier instantaneously at a non-nodal point is infinite. A slower insertion of the barrier, up to the adiabatic case, also requires work. Therefore, if one sees an experimenter `sweat' while inserting a barrier, then the tested wave function must have been probed at a non-nodal point.
This dramatically different behaviour from the insertion of a barrier at a nodal point suggests useful applications.
Next, we define a new test function 
\begin{eqnarray}
\chi_{before}(x)=\sqrt{\frac{1}{L}}\Big(\sin\Big( \frac{2\pi  x}{  L}\Big)+\sin\Big(\frac{ \pi  x }{ L}\Big)\Big)
\end{eqnarray}
with an initial transition probability with $\phi_{before}$ of $1/2$.
Since we know how each of the two components of $\chi_{before}$ reacts under an insertion, the linearity assumed in quantum mechanics allows one to immediately deduce how the new combined state $\chi_{before}$ is effected.
The overlap will not change with the insertion of the barrier and retains the value $1/2$. Does this suggest an inability to break the Helstrom bound, since the transition probability remains unchanged. This is not the case, because there is an additional source of information about the states given by the energy required to insert the barrier. This will be explored in the next section.


Before we exploit this result in the next section, let us state explicitly two potentially controversial assumptions.
 The first assumption is to rely on an insertion of an impenetrable barrier to divide the well.  This type of procedure was already discussed in different forms, i.e. from the instantaneous to the adiabatic, in an earlier paper by
Bender {\it et al.}  \cite{bbm} and seems not too unrealistic. 
The second assumption is that one is able to measure the energy needed to insert a barrier. A insertion at any speed from the instantaneous to the adiabatic would give a clue about the state one is probing, since energy is always needed to insert the barrier in a non-nodal point. The adiabatic case has  been discussed in the literature \cite{bbm} and in other papers by the same authors and leads to a more palatable  finite increase of the energy in the non-nodal case.
In the next section we use the preliminary results obtained so far to calculate the Bayes cost before and after the insertion.

\section{Calculation of the Bayes cost before and after the division of the chamber}

This section starts by restating the Bayesian approach to the binary decision problem as developed by Helstrom \cite{helstrom}. The result is applied to the cost calculation before and after the division of the square well.
The standard optimal lower bound in the binary decision problem with a $0-1$ cost, where cost $1$ is assigned to an incorrect decision and cost $0$ for a correct decision, is given by the Helstrom bound,
\begin{eqnarray}
&&C(\xi,|\bra\phi_{before}|\chi_{before}\ket|^2)\nonumber\\
&=&\frac{1}{2}-\frac{1}{2} \sqrt{1-4\xi(1-\xi)|\bra\phi_{before}|\chi_{before}\ket|^2},\nonumber
\end{eqnarray}
  for the two states $|\phi_{before}\ket$ and $|\chi_{before}\ket $, and their respective prior $\xi$ and $1-\xi$.   For a compact derivation, an example of a measurement setup for spin-1/2 particles to achieve this bound,   and  an extension  to the case of multiple copies see Brody {\it et al.} \cite{dbbm}.

In the following paragraphs the cost is evaluated before and after  the insertion.
Prior to the insertion the Helstrom bound for the two chosen states
\begin{eqnarray}
\phi_{before}(x)&= & \sqrt{\frac{2}{L}}\sin\Big( \frac{2\pi  x }{L}\Big), \nonumber\\
 \chi_{before}(x)&= &\sqrt{\frac{1}{L}}\Big(\sin\Big( \frac{2\pi  x}{  L}\Big)+\sin\Big(\frac{ \pi  x }{L}\Big)\Big) \nonumber
\end{eqnarray}is
\begin{eqnarray}
\frac{1}{2}-\frac{1}{2}\sqrt{1-2\xi(1-\xi)}\nonumber
\end{eqnarray}
with the transition probability equal to $1/2$.
This transition probability stays unchanged through the insertion of the barrier, since the new amplitude  of both wave functions except for a set of measure zero matches the original wave functions. 

A tentative estimation of the cost after the insertion follows next. The  probability for finding the particle in each of the new compartments after the insertion is equal for both  test wave functions. The overlap between the wave functions also remains $1/2$. As an aside, one can  construct a set of test functions so that the overlap on each side of the barrier matches the initial overlap, e.g. by choosing the two test functions 
to be   the first excited state and a combination of the first, second and fourth excited state with appropriately chosen weights (with the weight of the fourth excited state $-21/10$ times as a large as the weight of the second excited state).  
  
  After the insertion there are two sources of accessible information. The first is associated with the
new wave functions at the left and right side of the barrier. These new wave functions can be tested using conventional Helstrom strategies to reconfirm the standard Helstrom bound. 
There is a novel source of information obtained from the measurement of the energy needed to insert the barrier. This second and new type of information provides additional insight, since it differs markedly in the two cases.
If the first test function is in the well, the needed insertion energy is zero. If the second test function is in the well, a substantial amount of energy, infinite in the idealized instantaneous case, is necessary.
This additional information, even with an imprecise measurement of the insertion energy, should always reduce the cost below the Helstrom bound. The exact cost reduction depends on how precisely one can place the barrier, the speed of insertion, and the measurement uncertainty associated with the energy needed to insert the barrier.
If each of these three points can be addressed satisfactorily, then one can calculate the exact improvement of the optimal cost beyond the  Helstrom bound. In anycase, if the energy is significant, the probed state is likely to be the one with the non-nodal insertion.
This is the insight of the paper: Inserting a barrier requires energy for a non-nodal point, provides extraneous information, and can be used to help distinguish between different states.
%

\section{Conclusion}

It has been shown that the Helstrom bound in the binary quantum discrimination case can be breached, if the energy needed to insert a barrier can be measured with any precision.
An intuitive way to grasp the result is to link the additional information provided by the energy required to insert the barrier, i.e.
the insertion of a barrier into non-nodal is more energy intensive than the insertion at a nodal point, to the ability to distinguish the states.

 In a companion paper to be presented at an upcoming conference potential implications for quantum algorithm will be assessed. Almost without fail quantum algorithms can be viewed as procedures to distinguish between different states.  The method described above can be extended to states with arbitrary overlap.
 The important step is always to map the two or more candidate wave functions with arbitrary overlap into
  states with distinct nodal points without initially changing the distance between the candidate states. 
 Experimental implications, for example in the area of Bose-Einstein condensate, have not been studied, but should be of interest.
 

\hspace{-.38cm}The author wishes to express his gratitude to D.C. Brody for  stimulating discussions.

%

\end{document}